# Pixelated Antenna with Enhanced Isolation

Md. Amanath Ullah, *Graduate Student Member, IEEE,* Rasool Keshavarz, *Member, IEEE,* Justin Lipman, *Senior Member, IEEE,* Mehran Abolhasan, *Senior Member, IEEE* and Negin Shariati, *Member, IEEE*

*Abstract*—This paper presents a pixelated cubic antenna design with enhanced isolation and diverse radiation pattern for vehicular applications. The design consists of four radiating patches to take advantage of a nearly omnidirectional radiation pattern with enhanced isolation and high gain. The antenna system with four patches has been pixelated and optimized simultaneously to achieve desired performance and high isolation at 5.4 GHz band. The antenna achieved measured isolation of more than -34 dB between antenna elements. The overall isolation improvement obtained by the antenna is about 18 dB compared to a configuration using standard patch antennas. Moreover, isolation improvement is achieved through patch pixelization without additional resonators or elements. The antenna achieved up to 6.9 dB realized gain in each direction. Additionally, the cubic antenna system is equipped with an E-shaped GPS antenna to facilitate connectivity with GPS satellite. Finally, the antenna performance has been investigated using a simulation model of the vehicle roof and roof rack. The reflection coefficient, isolation and radiation patterns of the antenna remains unaffected. The antenna prototype has been fabricated on Rogers substrate and measured to verify the simulation results. The measured results correlate well with the simulation results. The proposed antenna features low-profile, simple design for ease of manufacture, good radiation characteristics with multidirectional property and high isolation, which are well-suited to vehicular applications in different environments.

## I. Introduction

THE use of wireless technology is ubiquitous in our everyday lives. After a few decades of focusing on the development of wireless systems, there is now increased attention to vehicular communications [1-3]. Communication between vehicles can be incorporated into a telemetric platform to supply the driver with real-time information. To provide a better user experience, mobile and wireless connectivity in vehicles is gaining increasing interest from academia and industry. A wireless local area network (WLAN) system for internet access from any location, as well as vehicle-to-vehicle communication systems for safe driving control could be included in future vehicles. Radio frequency (RF) systems such as Satellite Digital Audio Radio Service (SDARS), Global Positioning System (GPS), Vehicle to-Everything (V2X) communication, cellular communication, and Wireless Local Area Network (WLAN) are increasingly installed in modern automobiles for navigation, communication, and entertainment purposes [1, 4-7].

A vehicle's antenna design is challenged by stringent requirements. Vehicular communication modules require low-profile antennas with multi-standard characteristics, such as good radiation performance with the ability to communicate in multiple directions, enhanced isolation from antenna elements, high gain and high bandwidth. Increased coverage, enhanced robustness to multipath, and resistance to signal interception and interference are some of the benefits that smart antennas offer over standard antennas. Also, smart antennas have the ability to determine the direction of arrival (DOA) of an incoming signal and adjust the radiation pattern to enhance smart vehicular localization systems [8, 9].

A multidirectional antenna system is a promising approach to meet the requirements of vehicular technologies. There is a greater chance of improved signal reception if the antenna has more radiators. Because of this feature, the antenna can establish a reliable connection even when the vehicle is in motion. Simple omnidirectional patch antennas can be utilized for this purpose as omnidirectional antennas may receive signals from any direction (both in terms of elevation and azimuth), but their gain is typically quite low [10]. In such a scenario, the receiver may feature a number of antennas oriented in various directions to detect signals in its various polarizations. The quality of the received signal is enhanced by employing appropriate switching/selection or combining strategies.

In such a scenario, the issue of mutual coupling comes into effect when multiple antenna elements are closely placed together. Antennas should be as far apart as possible from one another to minimize interference from nearby radiators and maintain signal quality.

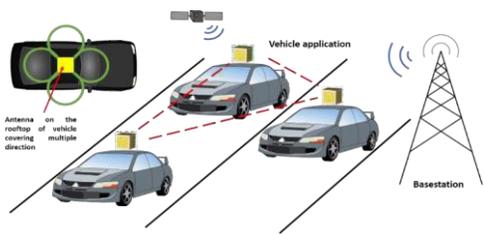

Fig. 1. Potential vehicular application scenario of the proposed multidirectional antenna in outdoor environment.

Therefore, the best way to establish a solid connection is through increased isolation [6, 11, 12]. Different attempts have been made by the antenna research community to improve isolation. The defected ground structure (DGS) is a common technique for reducing mutual coupling [13-15]. Nevertheless, antenna radiation performance is impacted when the DGS is used as a decoupling element. As a result, improved isolation comes at the expense of radiation performance. Considerable separation between the antenna elements is required to reduce mutual coupling using the split-ring resonators (SRR) [16] and electromagnetic bandgap (EBG) [17] structures [18]. In addition, other mutual coupling reduction techniques include slotted meander-line resonators (SMLR) [19], metamaterial structures [20] and parasitic elements [21] between antennas. However, such methods involve utilization of extra space within the antenna structure, as they are mostly used as separate elements or resonators to suppress mutual coupling. In addition to increasing the antenna size, they may affect the antenna performance.

As seen in Fig. 1, multidirectional antenna systems can facilitate real-time, traffic-based adjustment of the coverage area. High-efficiency and energy-saving communication is therefore possible. Vehicle antennas should be able to communicate in multiple directions to provide communication dependability between the vehicle and other wireless infrastructure, particularly in the case of long-distance, wide-area coverage, and high-speed data transmission.

In this paper, we propose a multidirectional cubic shaped antenna system with enhanced isolation and high gain. The structure consists of four antennas with pixelated configurations on the patches. The antenna design is based on binary optimization of the radiating patches while focusing on two design goals of reflection coefficient at the desired frequency (5.4 GHz) and isolation improvement. This work suggests combining polarization diversity measures to simultaneously take advantage of omnidirectional patterns with high gain. To increase the possibility of capturing wireless signals, multiple directional antennas are arranged in the cubic structure configuration. The antenna is designed using a binary optimization algorithm with pixelization of the radiating patches.

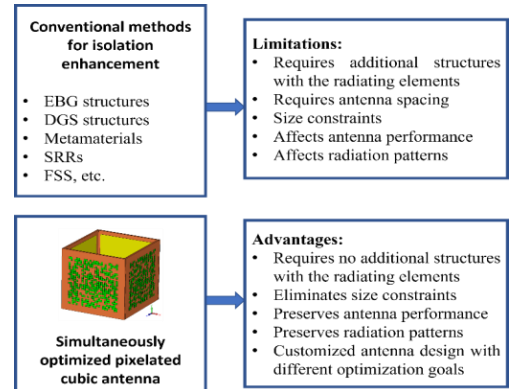

Fig. 2 Overview of the proposed antenna method: advantages over traditional design techniques to enhance isolation.

Exploration of different pixel configuration yielded desired results in previous dual-band antenna designs in [22, 23], where different kind of pixelated patch has been used to achieve operation in different frequency. To the best of authors' knowledge, the pixel optimization implemented to enhance isolation of multi antenna system in this work without using additional resonators has not been implemented in any prior work. No extra resonator or parasitic element is placed between the antennas, facilitating a compact antenna profile, while maintaining desired performance with increased isolation. Moreover, an edge-to-edge gap between the radiating elements is not required in this design. Fig. 2 illustrates the proposed antenna design advantages over conventional methods to enhance isolation. The proposed antenna achieved up to 6.9 dBi realized gain. The measured results demonstrate that the isolation between the antennas is better than −34 dB. The overall improvement of 18 dB isolation is achieved in comparison to a standard patch antenna system. Additionally, an E-shaped patch has been placed on top of the cubic structure, which operates at GPS frequency of 1.57 GHz.